\documentclass[runningheads]{llncs}
\usepackage[T1]{fontenc}
\usepackage{cite}
\usepackage{amsmath,amssymb,amsfonts}
\usepackage{graphicx}
\usepackage{textcomp}
\usepackage{xcolor}
\usepackage{xspace}
\usepackage{todonotes}
\usepackage{adjustbox}
\usepackage[title]{appendix}
\usepackage{hyperref}
\usepackage{float}
\usepackage{chngcntr}
\usepackage{wrapfig}
\usepackage{subcaption}

\newcommand{\grad}{\textit{GradCAM}\xspace}
\newcommand{\lrp}{\textit{LRP}\xspace}
\newcommand{\lime}{\textit{LIME}\xspace}
\newcommand{\shap}{\textit{SHAP}\xspace}
\newcommand{\intgrad}{\textit{Integrated Gradients}\xspace}
\newcommand{\cs}{\textit{Confidence Scores}\xspace}

\begin{document}
\title{Study on the Helpfulness of Explainable Artificial Intelligence (XAI)}
%
%
\author{Tobias Labarta\inst{1, 2}
\and Elizaveta Kulicheva\inst{2}
\and Ronja Froelian \inst{2}
\and  Christian Geißler\inst{2}
\and Xenia Melman\inst{2}
\and Julian von Klitzing\inst{1, 2}}

%

\institute{Fraunhofer Heinrich-Hertz-Institute, 10587 Berlin, Germany\footnote{We express our gratitude to Fraunhofer Heinrich-Hertz-Institute for financially supporting our work.}
 \and Technische Universität Berlin, 10587 Berlin, Germany}

\maketitle              

\begin{abstract}
Explainable Artificial Intelligence (XAI) is essential for building advanced machine learning-powered applications, especially in critical domains such as medical diagnostics or autonomous driving. Legal, business, and ethical requirements motivate using effective XAI, but the increasing number of different methods makes it challenging to pick the right ones. Further, as explanations are highly context-dependent, measuring the effectiveness of XAI methods without users can only reveal a limited amount of information, excluding human factors such as the ability to understand it. We propose to evaluate XAI methods via the user's ability to successfully perform a proxy task, designed such that a good performance is an indicator for the explanation to provide helpful information. In other words, we address the helpfulness of XAI for human decision-making. Further, a user study on state-of-the-art methods was conducted, showing differences in their ability to generate trust and skepticism and the ability to judge the rightfulness of an AI decision correctly. Based on the results, we highly recommend using and extending this approach for more objective-based human-centered user studies to measure XAI performance in an end-to-end fashion.

\keywords{Explainable Artificial Intelligence (XAI)  \and Goodness measures and evaluation  \and Human-Centered XAI \and XAI User-Study \and XAI Evaluation.}
\end{abstract}

\section{Introduction}
\label{sec:intro}
Society is facing the exponential application and exploitation of deep neural network-powered artificial intelligence (AI) in various areas of daily life. AI systems that make use of learned models to solve classification, segmentation, and transformation tasks for different input modalities such as images, video, text, and natural spoken language have shown or have the potential to outperform humans on specific tasks \cite{farhat2020deep, grigorescu2020survey}. Since they are used in domains such as mobility, energy management, finance, medical diagnosis, and in general health, security, and many further critical domains, risk management and reduction plays a crucial role in building safe AI systems, which is also increasingly required by recent regulatory advances \cite{gdpr, eu2022ethics}.

To analyze and validate AI systems despite their statistical and often non-deterministic process of creation and their complex structure\cite{hu2021model}, instead of analytical proofs, proxy methods such as empirical evaluations and explainable AI (XAI) are used.
Empirical evaluations using test datasets and calculating performance measures such as a confusion matrix, f-scores or AUC provide a narrow view of what the model learned. A vast amount of very different explainable AI methods have been proposed to understand, question, and investigate the models' inner workings and decisions, not just during testing but also while they are being applied in a real context. Choosing the right set of them for a specific application context is hard, especially if that context requires human operators to stay in control of the AI system. Thus, it is essential to develop methods for measuring and comparing the performance of XAI methods, especially when applied within specific contexts aimed at achieving distinct human-centered objectives. We term this "human task-related performance helpfulness." In this framework, "helpfulness" is defined as a quantifiable improvement in user performance on tasks that are aligned with the goals facilitated by the provision of explanations. In our experiment, the primary function of the XAI is to aid individuals in making informed decisions about their trust in the AI's results.

Measuring effects on users is usually tested in a user study. Most of these studies focus on usability aspects such as satisfaction which are biased by user preferences and opinion. They are not general proof if a certain explanation has a positive effect on the safety-related performances of an AI system. They therefore measure trust (belief) related performance, not the ability to which the explanation helps to control the AI system.

In this paper, we address the following research questions:

\begin{enumerate}
\vspace{-2mm}
\item How accurate can XAI methods enable a user to \textit{judge} AI decisions?
\item How far can XAI methods enable a user to \textit{trust} AI decisions?
\item How far can XAI methods enable a user to \textit{question} AI decisions?
\vspace{-2mm}
\end{enumerate}

The rest of the paper is structured as follows: We start with an overview of the state of the art of measuring the performance of explainable AI methods. We continue by contributing a new approach for measuring XAI performance in human-centered user studies on objective performance criteria. We further demonstrate the approach by conducting a user study to compare six well-known XAI methods concerning their ability to enable the user to correctly judge the trustworthiness of an AI-based classification decision. Finally, we discuss the insights drawn by this human-centered, objective evaluation approach.

\section{Measuring Explainability}

\subsection{Approaches for measuring explainability}
Previous theoretical work on explainability and its measurement are often inspired by research from domains such as human-computer interaction, psychology, philosophy, or machine learning. A common finding in most works is, that explanations are context dependent. Therefore, they are required to define the recipients of the explanation (the explainees), the use case in which they operate the AI system, and the specific situation in which the explanation is provided \cite{liao2022connecting, carli2022risk, van2021evaluating}. The typology of an explanation is given by an explanandum, the object (thing or phenomenon) to be explained, the explanans (the actual explanation that is perceived by the users), and the relationship between both \cite{cabitza2023quod}. Cabitza further provides the following broad definition for the explanation provided by an XAI-system: It is \textit{"the output of any computational system aimed at making AI-generated advice more understandable, appropriable and exploitable by their intended users and decision makers"}.

In contrast to this specific perspective, some suggest measuring the degree of completeness of an explanation by providing a set of all possible questions \cite{liao2020questioning, salewski2022clevr, sovrano2022quantify}. "A set of all possible questions" refers to a complete collection of questions that could be asked to evaluate and understand an AI's behavior and decisions fully. Effectively, this would covers all aspects of the AI's operation, rationale, outcomes, and underlying mechanisms to ensure that the explanation of the AI is complete. The idea behind providing such a set is to measure how well an explanation satisfies the informational needs of users concerning the AI system. If an explanation can address all possible questions, it can be assumed complete in the sense that it leaves no aspect of the AI unexplained. However, providing such a complete set of questions for a real, practical use case is challenging as it requires interpretation in the form of a transfer of the questions into the specific application context. This introduces subjective variability and therefore, makes it difficult to objectively compare the explanatory power between different contexts. However, having a set of template questions can support formulating questions a user might come up with when designing an AI system. This is further supported by works like \cite{liao2022connecting}, which investigate the importance of such different questions and criteria. The results of that work highlight that faithfulness and translucence are the most important criteria and that experts and end-users mostly align in how they rate each criterion.

Besides the local or global perspective, measuring explainability can also be categorized into human-centered (e.g. that require human participation via a user study) and methods without human involvement \cite{vilone2021notions}. Approaches without human feedback, for example, because they use an artificial benchmark with defined ground truth for the desired explanations \cite{muller2021kandinsky, salewski2022clevr, hedstrom2022quantus}, suffer from a direct proof of being transferable to real-world use cases. For example, since explanations are context-dependent, it is not evident that they transfer to other contexts. However, they are significantly cheaper to apply than user studies. 

Human-centered evaluations often measure subjective qualities such as user satisfaction or opinion by directly questioning participants. While they provide a holistic perspective on the explanation as they are embedded in the AI system and context, they also introduce additional noise. They do not directly measure the desired qualities but have to rely on proxy questions \cite{vilone2021notions} which introduce human bias \cite{bertrand2022cognitive}.

With our study, we want to extend and motivate human-centered evaluations to measure not just satisfaction, but also the degree to which the explanans helps the users to reach their goals in terms of objective task performance.

\subsection{User studies on the performance of XAI} 

Within the health domain, a study on XAI was conducted ~\cite{evans2022explainability} to evaluate the helpfulness of various approaches to AI assistance in digital pathology. Clinical pathologists were shown various examples of the AI-assistance in a questionnaire, combined with expert interviews. Results of the study show a preference for pathologists for simple visual explanations that correspond to their way of making diagnostical decisions. On the other hand, participants expressed a concern that simplistic explanations allow for a lot of ambiguity in their interpretation. 

A study on XAI in the automotive domain~\cite{kim2018textual} investigated the difference in perception of the end users of seeing a textual description of the decision and seeing a textual explanation of the decision without a decision itself. A user was asked to evaluate whether the AI model made a correct decision based on a textual description or an explanation of said decision. Results of this survey show that users tended to trust the explanation without a decision rather than a decision without an explanation. 

Lakkaraju and Bastani \cite{lakkaraju2020fool} performed a study on the impact of misleading explanations on users. They hypothesized that existing XAI trust measures are not sufficient, as explanations could be perturbed, leading to users trusting a problematic AI. For this purpose, they constructed a black-box AI making bail decisions based on prohibited features like race or gender. On top, they added an intrinsic explanation method but distorted it such that it returns other, desired features as explanations that were not part of the decision, e.g. prior jail incarcerations.  After conducting the study with 41 participants, they could show that users were 9.8 times more likely to trust the black-box AI with a misleading explanation than just the black-box AI although the AI was making the same, wrong decisions in both groups. They advocate for more interactive and explorable explanations, which they back with research findings \cite{lakkaraju2019faithful} and a second, slightly adapted study they conducted.

Ribiero et al. \cite{ribeiro2016should} present an explanation method that is evaluated in this study, \lime. In addition to that, they conduct two studies to evaluate their newly developed method. First, they perform simulated user experiments to investigate the explanation usefulness of \textit{LIME} and compare it to other methods. Second, they evaluate \textit{LIME}  with real participants. Here, they checked if users can decide which model performs better, based on the explanations. Ribiero et al. also asked the users whether they trust the decision of a biased model, in this case, the "Husky vs. Wolf" example \cite{murdoch2019definitions,hodges2019machine,badillo2020introduction}. They could show, that after displaying an explanation that highlights the model bias, significantly fewer users trusted the bad model.

A different approach presented by~\cite{matarese2023much} is to measure and quantify the amount of information the model provides, by measuring the number of rules a user discovers while interacting with an XAI system. While great at providing concrete and quantitative numbers the experiment setup, relies heavily on human interpretable features, falling short for complex deep learning models.

A recently published user study is by Achtibat et al. \cite{achtibat2023attribution} where they explored the practical utility of Concept Relevance Propagation (CRP), a promising development from Layerwise Relevance Propagation. Using provided explanations as guidance, users had to determine if the model’s prediction was affected by a bias. In this scenario, two image classification models were used, where one was trained to take advantage of image borders for prediction. For both models, explanations were generated for CRP, but also for other popular XAI methods, e.g. SHAP, Grad-CAM, etc.,  that were also used in our study. As the design of this user study allowed for objectively true or false replies from the participants, the authors applied common classification result tools such as the confusion matrix and calculated accuracy measures for each method.

\section{An objective Methodology for evaluating XAI}
\label{sec:methodology}

\begin{figure}[t]
\centering
\includegraphics[width=0.9\textwidth]{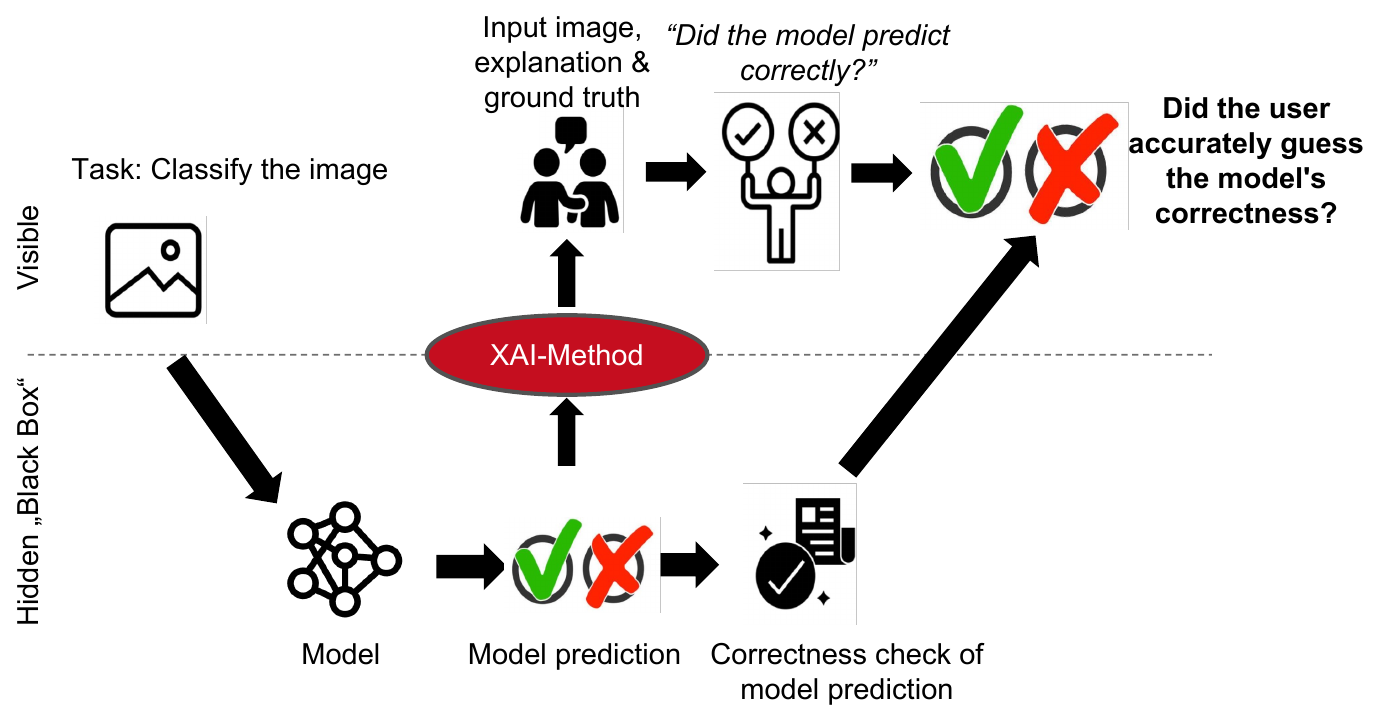}
\caption{Overview of the objective methodology for evaluating XAI. It starts with the classification of an image. The model makes a prediction hidden from the user, that will be checked by its correctness in the background. Then, a local post-hoc XAI method is applied to generate an explanation that will be presented to the user, together with the input image and ground truth. Based on this information, the user has to predict the model prediction.} 
\label{fig:experiment_flow}
\vspace{-5mm}
\end{figure} 

A key finding from the literature research was the lack of objective methods for evaluating XAI methods in a human-centered setting. Most of the existing approaches focus on qualitative studies investigating subjective features like satisfaction or just collecting feedback. Quantitative Methods without human involvement are often called to be more objective but are lacking the user focus which is essential for applied XAI methods. The only approach combining human involvement with quantitative methods was by Achtibat et al. \cite{achtibat2023attribution}, where users were asked to assess if the model relied on artifacts in the prediction process. This approach is a very interesting step in the right direction for quantitative XAI user-studies, but also focused on a sub-task, namely detection of model bias. To allow for a more general approach, our work proposes an objective methodology for human-centered evaluations of XAI methods and executes it on six commonly used explainability methods. An overview of this novel methodology is shown in Figure \ref{fig:experiment_flow}.

\subsection{Objective Human-Centered XAI Evaluation}
\label{sec:exSetup}
We propose to use a proxy task, to evaluate our approach and answer the research questions about the ability of XAI methods to enable the user to judge, trust, and question AI decisions. The task is designed such that the performance to solve it directly represents the ability to use the explanations for the investigated purpose, e.g. in our case, to determine if the explanation allows the user to judge whether a model predicted the class of an input image correctly. Participants should judge this by reviewing the input image, knowing the ground truth label and the output of a single explainability method. During the experiment, the participants were not aware of the model's actual prediction or its correctness. An example of this setup can be found in Fig.~\ref{fig:experiment_flow}.

\subsubsection{Evaluation Metrics}
\label{sec:evalSetup}
As participants judge whether a model's output is correct, their prediction can either be correct or false. Therefore, their replies can be viewed as a confusion matrix of a binary classifier \cite{yerushalmy1947statistical} as summarized in table \ref{table:confusion_matrix}. With this abstraction, it is possible to calculate participants' accuracy, sensitivity, and specificity.
 \textit{Accuracy} $=\frac{TP+TN}{TP+TN+FP+TN}$  expresses the participant's ability to correctly \textit{judge} model predictions. As we used a balanced number of instances, an accuracy of 0.5 is the baseline for random guessing. This baseline means that in a binary classification task in balanced settings, the simplest random guessing will result in correct judgments about half the time, assuming an equal likelihood of either outcome. This sets a baseline accuracy of 0.5 as the point of comparison, indicating that any performance above this threshold suggests a better-than-random ability to discern the correctness of the model's outputs.
\textit{Sensitivity} and \textit{specificity} are used to measure participants' ability to identify when they can \textit{trust} a model's prediction and when they should \textit{question} a model's prediction. The \textit{sensitivity} is also known as true positive rate ${TPR} = \frac{TP}{TP+FP}$. The \textit{specificity} is also known as the true negative rate ${TNR} = \frac{TN}{TN+FN}$.
\newline
\begin{table}[H]
\vspace{-15mm}
\caption{Confusion matrix}
\label{table:confusion_matrix}
\begin{center}
\begin{adjustbox}{width=1.0\textwidth}
\begin{tabular}{| c | c | c |} 
 \hline
 & User assumes the model is correct & User assumes the model is wrong \\
 \hline
model output was correct & True Positive (TP) & False Negative (FN) \\
 \hline
model output was false & False Positive (FP) & True Negative (TN) \\
 \hline
\end{tabular}
\end{adjustbox}
\end{center}
\end{table}
\subsubsection{Hypotheses testing and effect size}
\label{sec:hypo_test}
To check if the survey results also hold for the general population, hypothesis testing was applied. The significance level $\alpha$ was set to $0.05$ for all tests.\newline
A \textit{one-sample t-test} was performed per method. This test is supposed to verify if a sample metric mean of a method can be considered significantly larger or lower than the random baseline. Thus, the null hypotheses ($H0$) and alternative hypotheses ($HA$) were set according to the resulting sample metric mean size. 
To measure the effect size of a method in comparison to the random baseline, the effect size \textit{Cohen's} $d$\cite{cohen2013statistical} for the \textit{one-sample t-test} was calculated. The measure is only calculated if the corresponding test result is significant.
To evaluate which XAI methods perform best a \textit{paired t-test} was performed for each pairwise method combination. 

\subsection{Image classification \& XAI methods}
\vspace{-8mm}
\begin{figure}[h]
    \centering
    \includegraphics[width=0.65\textwidth]{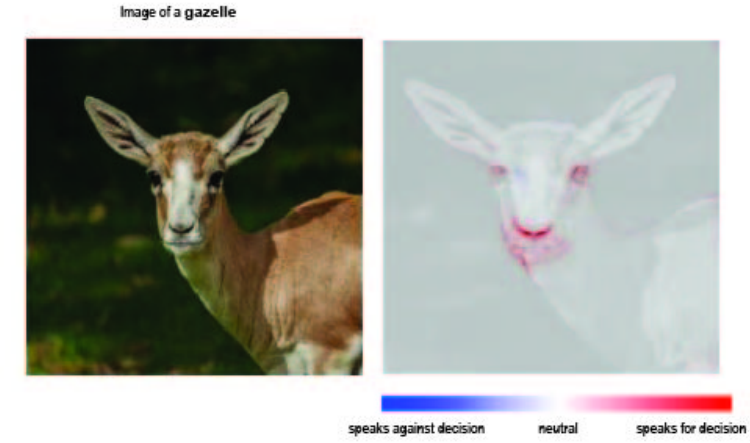}
  \caption{Example Survey Question}
\label{fig:question_example}
\vspace{-7mm}
\end{figure}
To avoid model-specific biases, two different AI models, AlexNet~\cite{krizhevsky2017imagenet} and VGG16~\cite{simonyan2014very}, were selected for classification. The models were used without hyperparameter optimization. As image source, the \textit{imagenetv2-matched-frequency} dataset~\cite{imagenetv2} was chosen. It is a subset of \textit{imagenet}, which originally contains 1000 classes, and each class has 500 images on average. The objects shown in the images range from everyday items to specific animals. \newline

Based on the experimental setup with a classification of individual images, only local explanation methods fit for the imaging modality could be used. Additionally, the methods had to be applied post-hoc so they could work on the same set of images. Other selection criteria included the use of well-documented methods and the availability of code. Acceptance and relevance within the scientific community were also key factors in the selection process. Following these restrictions, six XAI methods were selected: \textit{Layer-wise relevance propagation} (\lrp)~\cite{bach2015pixel,montavon2019layer}, \textit{Gradient class activation mapping} (\grad)~\cite{chattopadhay2018grad, das2020opportunities}, \textit{Local Interpretable Model-Agnostic Explanations}  (\lime)~\cite{garreau2021does}, \textit{SHapley Additive exPlanations} (\shap)~\cite{speith2022review,lundberg2020local,lundberg2017unified,shapley1997value}, \textit{Integrated Gradients} (\intgrad)~\cite{sundararajan2017axiomatic} and \textit{Confidence Scores}. \newline
\cs is the only numerical explanation type used in this study. It provides information on how confident the AI model is that the prediction with first, second, or third rank is correct. It also provides information about the confidence gaps between the ranks. 
The confidence score is generated based on the normalized results of the last model layer before picking the highest one as the one the model outputs as a prediction. Therefore, it is not a strict probability that the result is correct, but rather a comparative value about how close the highest-ranked decisions were.

\subsection{Survey design}
\label{chap:survey_design}

Besides a set of demographic questions, incl. educational background, machine learning experience, and visual impairment, the survey consisted of 12 independent questionnaires with 24 pictures each. Of these 24 pictures, 12 were the same across all questionnaires to provide a reliable baseline and the other 12 were semi-randomly picked. They were picked to ensure that each combination of the XAI method, AI model, and output was uniformly represented. Details of this process can be found online\footnote{complete link : \href{https://github.com/tlabarta/helpfulnessofxai}{\color{blue}https://github.com/tlabarta/helpfulnessofxai}}.

To minimize biases, the following precautions were made:
All XAI output was used to generate a heatmap which then was overlayed over the original image. Additionally, the heatmaps used the same colormap, one suitable for most color-based visual impairments, and a color bar was added to make the survey as accessible as possible. For reference see Fig.~\ref{fig:question_example}.

\section{Survey Results}
\subsection{Questionnaire responses}
\label{sec:quest_response}
\begin{figure}[h]
\centering
\vspace{-3mm}
\begin{subfigure}{1\textwidth}
    \centering
    \includegraphics[width=0.75\textwidth]{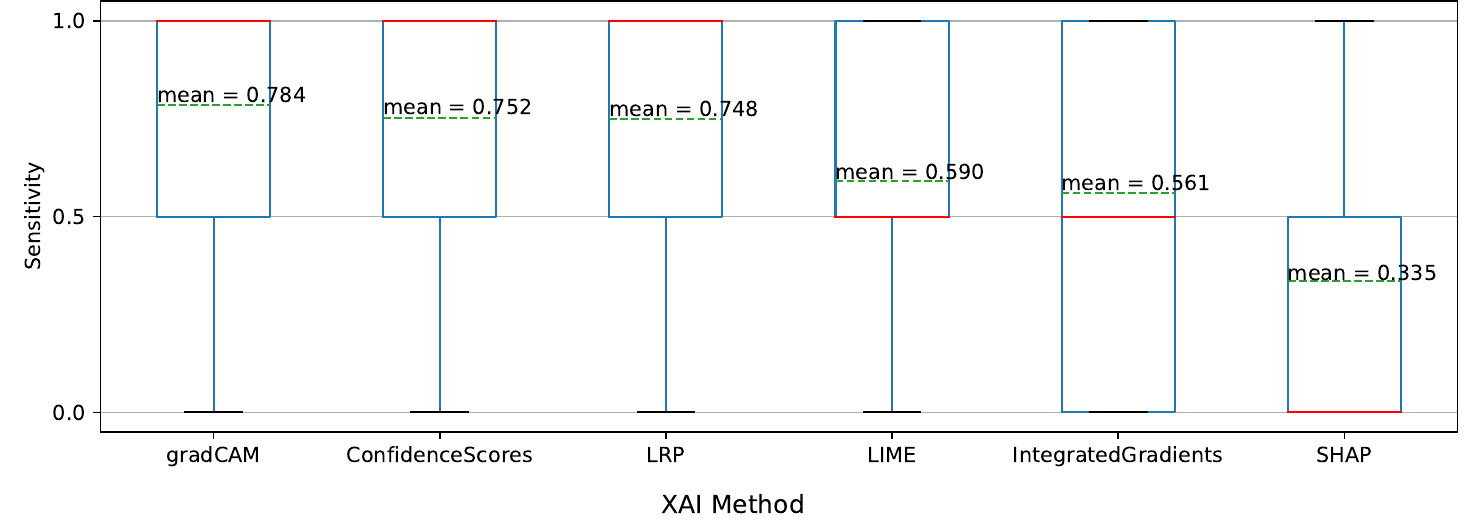}
    \caption{sensitivity}
    \label{fig:box_sensitivity}
\end{subfigure}
\begin{subfigure}{1\textwidth}
    \centering
    \includegraphics[width=0.75\textwidth]{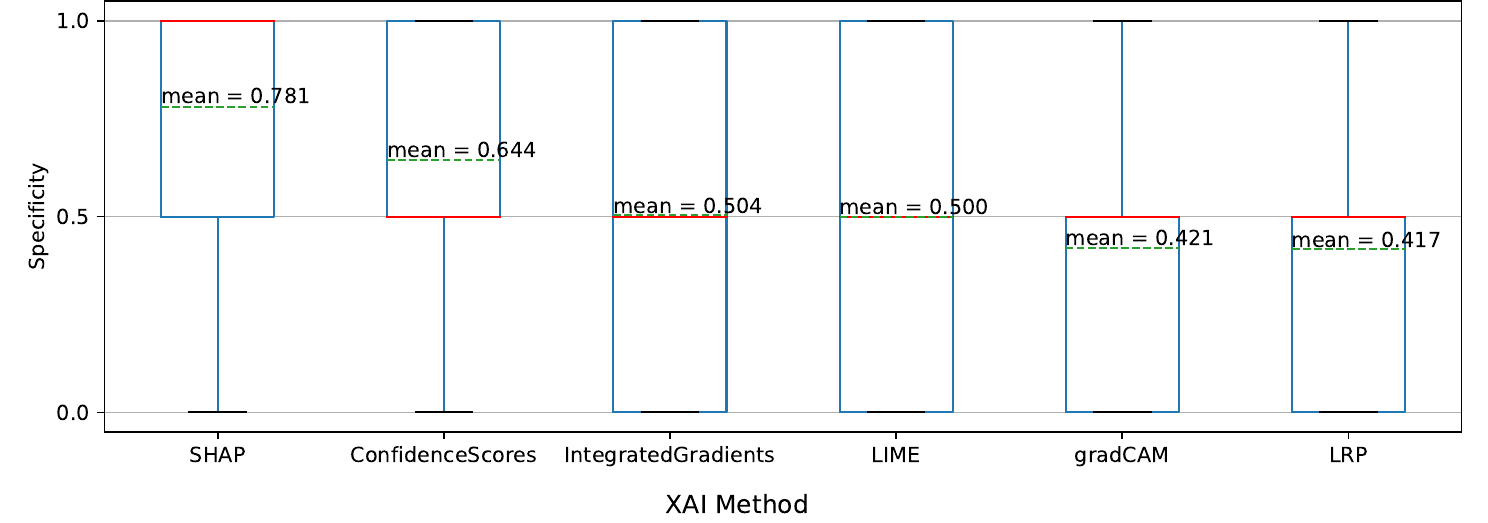}
   \caption{specificity}
    \label{fig:box_specificity}
\end{subfigure}
\caption{Participants' mean sensitivity \& specificity in judging whether the AI was correct in its prediction, per XAI method. SHAP performs best for sensitivity and worst of all methods for specificity. Confidence Scores perform the best in both measures, besides SHAP.} 
\label{fig:box_sens_specs}
\vspace{-0.2mm}
\end{figure}

The survey was closed after one month of execution time. Until the date of 13.07.2022, 139 participants completed the questionnaire. The survey was advertised using the TU Berlin social media pages, resulting in a substantial number of participants with a university background. Of all participants, 24 were undergraduates, while 44 were graduate students. Additionally, 26 postgraduate students took part, as well as seven participants with a PhD. Among the participants, no bias due to education, experience with machine learning, or visual impairment could be identified (see Appendix A for mean accuracy based on demographic groups). A more detailed demographic overview of the participants can be found in Appendix B.
To illustrate the results, the \textit{accuracy}, \textit{sensitivity}, and \textit{specificity} results of the 139 participants are shown. The accuracy convergence over an increasing number of participants can be found in the Appendix, see \ref{fig:res_coner}.
Figure \ref{fig:box_accuracy} presents the \textit{accuracy} results. Due to the questionnaire generation procedure, each participant could answer four questions per XAI method. Thus, the only possible results per XAI method were $0.0$, $0.25$, $0.5$, $0.75$, and $1.0$. The highest mean \textit{accuracy} over all participants was achieved by \textit{Confidence Scores} with $\approx$ 0.698. This was followed by \textit{GradCAM} $\approx$ 0.603, \textit{LRP} $\approx$ 0.583, \textit{SHAP} $\approx$ 0.558, \textit{LIME} $\approx$ 0.545 and \textit{Integrated Gradients} $\approx$ 0.532. Except from \textit{Confidence Scores}, which reached a median of 0.75, all other methods had a median of 0.5.
Figure \ref{fig:box_sensitivity} shows two opposite metrics: the \textit{sensitivity}, as well as the \textit{specificity} results. Based on the questionnaire generation procedure, each participant could answer two questions per XAI methods where the models decided \textit{correctly}. Thus, the only possible \textit{sensitivity} results per participant were $0.0$, $0.5$, and $1.0$. The highest mean over all participants was achieved by \textit{GradCAM} with $\approx$ 0.784, followed by \textit{Confidence Scores} with $\approx$ 0.752, \textit{LRP} $\approx$ 0.748, \textit{LIME} $\approx$ 0.590, \textit{Integrated Gradients} $\approx$ 0.561 and \textit{SHAP} $\approx$ 0.335. \textit{GradCAM}, \textit{Confidence Scores} and \textit{LRP} reached a median of 1.0, \textit{LIME} and \textit{Integrated Gradients} a median of 0.5, whereas \textit{SHAP} reached a median of 0.0. Due to the chosen questionnaire generation procedure, each participant could answer two questions per XAI method where the chosen models decided \textit{incorrectly}. Thus, the only potential \textit{specificity} values per participant were $0.0$, $0.5$, and $1.0$. The highest mean over all participants was achieved by \textit{SHAP} with $\approx$ 0.781, followed by \textit{Confidence Scores} with $\approx$ 0.644. \textit{Integrated Gradients} reached $\approx$ 0.504, \textit{LIME} = 0.5, \textit{GradCAM} $\approx$ 0.421 and \textit{LRP} $\approx$ 0.417. \textit{SHAP} reached a median of 1.0 whereas the other XAI methods had a median of 0.5.
Participant responses could be influenced by the model that an explanation was generated for. 

As described in section \ref{chap:survey_design}, the predictions were split half/half between both models. Since VGG16 generally achieves a better task performance than AlexNet one would assume a noticeable impact on the participant accuracy across all XAI methods. The assumption was that a better-performing model leads to more profound decisions and weights, which would have a positive impact on the generated explanations. Figure \ref{fig:res_models} in the Appendix shows that only a small difference between the two models existed. Participants performed marginally better when explanations were generated on VGG16 predictions, with an accuracy of 0.61 versus 0.56 for explanations generated on AlexNet predictions. There is little to no difference in participant performance between the two models for \intgrad{}, \shap{} and \grad{}. A noticeable difference can be seen for \cs{}, \lime{}, and \lrp{}.

\subsection{Qualitative Feedback}
In addition to the questionnaire responses, participant feedback about the study was received. Some of the feedback could be valuable for future studies.
One of the participants pointed out that it was hard to decide without knowing if a model was trained to recognize a specific class. On one hand, it would help the participant to make a decision, but on the other hand, it would also enable certain bias since it is possible to assume a participant would rather tend to answer "yes" to those classes which are in the trained classes list, and rather "no" to those classes not on the list. Also due to the broad participant scope of this study, it can be assumed that this information would not be helpful for all participants.
Another point was in the case of multiple objects shown in a picture. Feedback was that it would be helpful to highlight the object to be classified if there is more than one. For example, when having multiple dog breeds on an image, it would be good to know which one is to be classified. A disadvantage of this approach is that the model itself is not "told" which object it is asked to identify, this task belongs to the challenge of image recognition as well.
Another feedback was, that even people with a machine learning background had in some cases a hard time making a decision. This could explain the observed very low-performance differences between machine learning experts and non-experts.
\begin{figure}[h]
\vspace{-3mm}
\centering
\includegraphics[width=0.75\textwidth]{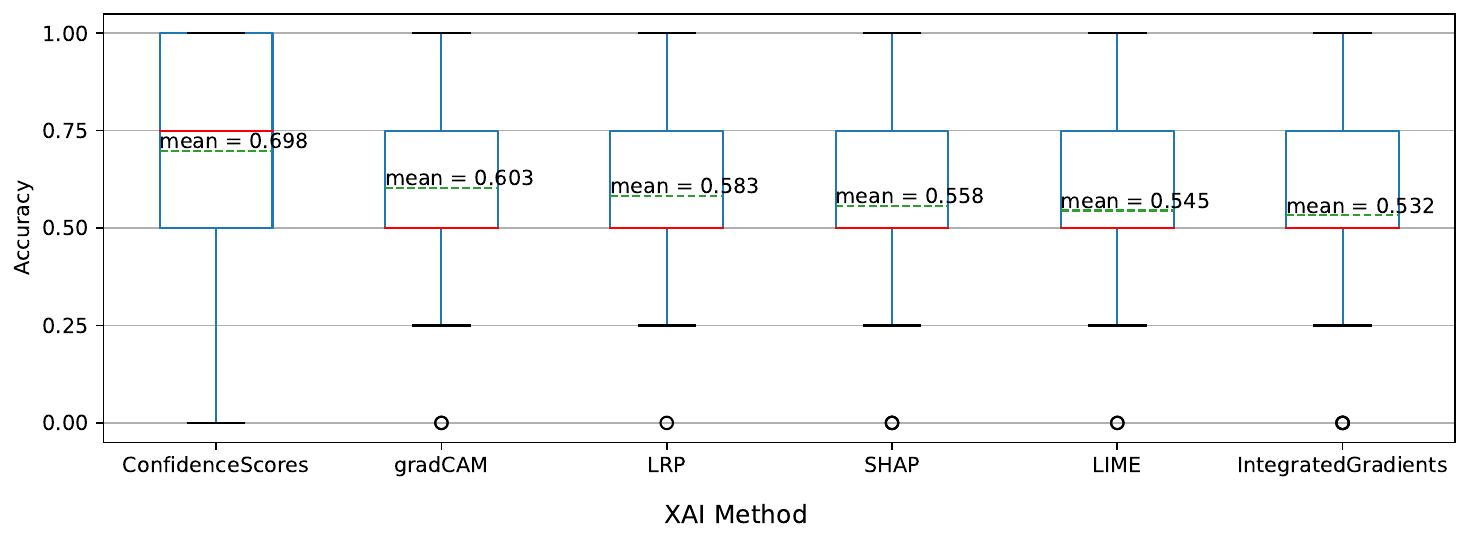}
\caption{The mean accuracy of participants in determining the correctness of the AI's predictions varied by XAI method used. Confidence Scores consistently outperformed all other methods, while the remaining methods showed minimal to no differences in results.} 
\label{fig:box_accuracy}
\vspace{-10mm}
\end{figure} 

\section{Discussion}

Continuing with the discussion, table \ref{table:Accuracy_table} states the results for the \textit{one-sample t-test} performed on the accuracy means. All means were significantly greater than the random baseline of 0.5 (better than chance). Only \textit{Confidence Scores}'s accuracy was significantly higher than for all other methods. These findings were drawn from the pairwise conducted \textit{paired t-test}s. 
Thus, \textit{Confidence Scores} helped a user most to \textit{judge} an AI's decision with a strong effect. All other methods were similarly helpful, with small to no practically measurable effect. \newline
It can be assumed that \textit{Confidence Scores}'s superior \textit{accuracy} (i.e. overall performance) originates partially from its relatively simple structure and clear results of just class probabilities. The explanations generated by the other methods have considerably more information than just classes with a probability and therefore need more interpretation. Participants had to consider the location of highlighted areas, as well as the color and brightness.

\begin{table}
\vspace{-6mm}
\caption{Hypothesis testing: accuracy}
\label{table:Accuracy_table}
\begin{center}
\begin{adjustbox}{width=.6\textwidth}
\begin{tabular}{||c|| c c c c ||} 
 \hline
 \textbf{XAI method} & \textbf{Accuracy} & \textbf{$HA$} & \textbf{\textit{p}} & \textbf{\textit{d}} \\ [0.5ex] 
 \hline\hline
 \cs{} & $\approx$ 0.698 & $>$ 0.5 & $<$ .001 & 0.84 \\
 \hline
 \grad{} & $\approx$ 0.603 & $>$ 0.5 & $<$ .001 & 0.46 \\
 \hline
 \lrp{} & $\approx$ 0.583 & $>$ 0.5 & $<$ .001 & 0.40 \\
 \hline
 \shap{} & $\approx$ 0.558 & $>$ 0.5 & $<$ .001 & 0.28 \\
 \hline
 \lime{} & $\approx$ 0.545 & $>$ 0.5 & .009 & 0.20 \\
 \hline
 \intgrad{} & $\approx$ 0.532 & $>$ 0.5 & .04 & 0.15\\
 \hline
\end{tabular}
\end{adjustbox}
\end{center}
\vspace{-5mm}
\end{table}

Furthermore, table \ref{tab:sens_spec_table} states the results for the \textit{one-sample t-test} performed on the \textit{sensitivity} and \textit{specificity} means. All XAI methods but \textit{SHAP} had means that were significantly greater than 0.5. However, only \textit{GradCAM} had a large positive effect. 

Overall, \textit{GradCAM}, \textit{Confidence Scores} and \textit{LRP} helped a user most to \textit{trust} an AI's decision with a medium to strong effect. Interestingly, \textit{SHAP} even negatively impacted trust in an AI's decision with a medium negative effect.
\begin{table}
\vspace{-6mm}
\caption{Hypothesis testing: specificity \& sensitivity}
\begin{center}
\begin{adjustbox}{width=1\textwidth}
\begin{tabular}{||c|| c c c c || c c c c ||} 
 \hline
 \textbf{XAI method} & \textbf{Specificity} & \textbf{$HA$} & \textbf{\textit{p}} & \textbf{\textit{d}} & \textbf{Sensitivity} & \textbf{$HA$} & \textbf{\textit{p}} & \textbf{\textit{d}} \\ [0.5ex] 
 \hline\hline
 \shap & $\approx$ 0.781 & $>$ 0.5 & $<$ .001 & 0.85  & $\approx$ 0.335 &  $<$ 0.5 & $<$ .001 & -0.43\\ 
 \hline
 \cs & $\approx$ 0.644 & $>$ 0.5 & $<$ .001 & 0.44 & $\approx$ 0.752 &  $>$ 0.5 & $<$ .001 & 0.77\\
 \hline
 \intgrad & $\approx$ 0.504 & $>$ 0.5 & .458 & not significant & $\approx$ 0.561 &  $>$ 0.5 & .037 & 0.15 \\
 \hline
 \lime & 0.5 & $\neq$ 0.5 & 1.0 & not significant & $\approx$ 0.590 &  $>$ 0.5 & .004 & 0.23\\
 \hline
 \grad & $\approx$ 0.421 & $<$ 0.5 & .007 & -0.21 & $\approx$ 0.784 & $>$ 0.5 & $<$ .001 & 0.89 \\
 \hline
 \lrp & $\approx$ 0.417 & $<$ 0.5 & .002 & -0.25 & $\approx$ 0.748 &  $>$ 0.5 & $<$ .001 & 0.74\\
 \hline
\end{tabular}
\end{adjustbox}
\end{center}
\label{tab:sens_spec_table}
\vspace{-5mm}
\end{table}
 \textit{SHAP} and \textit{Confidence Scores} achieved means that were significantly greater than 0.5. \textit{Integrated Gradients}'s and \textit{LIME}'s means were not significantly different from 0.5. \textit{GradCAM} and \textit{LRP} means were significantly smaller than 0.5. 
Thus, \textit{SHAP} helped a user most to \textit{question} an AI's decision with a strong effect.

The Imagenet dataset was chosen because of the assumption, that most people naturally are domain experts in classifying its images, as it shows objects of daily life. However, some participants mentioned that they had never seen some of the fruits, animals, or other objects that were presented before, meaning they lacked the required domain knowledge. This naturally impacts the ability to interpret the explanation, such as not being aware of the anatomic unique features of certain dog breeds.\newline
Usually, XAI methods are applied to help domain experts in evaluating model predictions. These should be aware of relevant details within their domain and their response might vary compared to our survey. In most applications, the field of predictions is much leaner and more focused on a specific application than the 1000 classes presented in ImageNetV2. 
\vspace{-5mm}
\begin{figure}[H]
    \centering
    \includegraphics[width=0.5\textwidth]{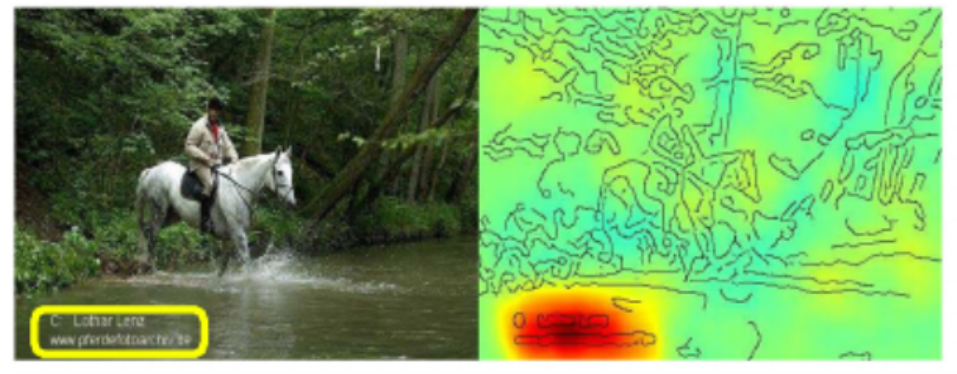}
  \caption{Right prediction based on a wrong feature}
  \label{fig:horse_example}
  \vspace{-7mm}
\end{figure}
Another critical point is the potential effect of so-called model bias on the survey results. It occurs in cases when a model makes a correct prediction based on wrong features. An example of this phenomenon can be seen in Fig.~\ref{fig:horse_example} where a horse picture from Pascal VOC data set was classified correctly by Fisher vector classifier~\cite{lapuschkin2019unmasking} because of a photographer tag in the bottom left corner. By seeing the explanation of this example, it should be possible for most users to detect the model bias and conclude that the model did not classify this image correctly. Detecting such model bias usually requires expert knowledge in the machine learning field. \newline
Noteworthy is also a potential bias in the study population, as it was shared within university context for everyone to attend, without demographic selection criteria. When analyzing participant performance based on demographic criteria, no statistically significant impact could be identified, however, that does not exclude potential bias from other demographic criteria that were not tracked.

Since a model made the right classification, even though based on completely irrelevant features, user decisions would contribute to false-negative samples in a confusion matrix. In the context of this survey, it reduces the accuracy and sensitivity of the XAI method. The problem here is that even though an XAI method showed a good explanation of the decision, its performance would still be degraded since the model did a bad job of learning the class features. This is a potential issue that could not be fully excluded from the survey, as there was no documented list of identified model biases available for VGG16 or AlexNet on ImageNet. \newline
To summarize, the results showed a clear superiority of the heuristic explanation method \cs{} in \textit{accuracy} and a good performance in \textit{sensitivity} and \textit{specificity}. A potential risk of the method is that purely heuristic explanations can lead to heuristic biases as they tend to oversimplify complex situations \cite{wang2019designing}. This oversimplification and an high enough trust on the users side can lead to over reliance~\cite{du2022role}, in which the users trust the model to much and let it cloud their judgement. A good explanation method needs to account for this by clearly visualizing the relevant features and combinations, to make it as easy as possible for the user to identify wrong predictions. In terms of \textit{specificity}, \textit{SHAP} was significantly above random baseline and superior to all other methods although it performed very badly for \textit{sensitivity} and slightly above random baseline for \textit{accuracy}. All in all, no explanation method overperformed in all cases. \newline
The results indicate that it would therefore be recommended to not rely on a single, monolithic explanation approach. Instead, a more diverse and interactive framework of explanation methods could be beneficial. This explanation framework should be focused on the explanation task and the user’s needs. Diverse in this context is understood as a selection of different explanation approaches rather than a single method. Such a framework could also help with reducing the previously discussed issues of the required domain and expert knowledge for certain explanation cases and could be more effective in detecting model bias. The design of such an explanation framework, as well as its evaluation, would be a research topic for the future.

A key contribution of this work was the proposal of a new, objective methodology for evaluating XAI. The survey executed with this methodology shows the benefit of the approach, extending highly subjective aspects of human-centered XAI evaluation with task performance-related ones. A point of critique is the relatively staged scenario of users not knowing the model decision. Usually, users should be aware of the decision but might not be able to judge whether it is correct or not. A thorough examination of the objective methodology as well as an approach that is closer to reality should also be part of future research.
\section{Conclusion}
Within this paper, an objective methodology for evaluating XAI methods was proposed. From literature research, a lack of quantitative methods for human-centered XAI evaluation was identified, which this work aimed to contribute to. This approach was evaluated in a user study, where six state-of-the-art XAI methods were implemented. The goal of the user study was to examine how far existing methods enable users to judge, trust, and question AI decisions.

The results show that of the tested methods only \cs{} substantially enabled users to judge an AI decision, responding to the first research question. Aimed at the second research question, the methods \grad{}, \cs{} and \lrp{} performed best in making users trust an AI decision. For the final research question, it can be concluded that \shap{} enabled users by far the most in questioning an AI decision.

From the research findings but also literature research, it can be concluded that using individual explanation methods is not sufficient for enabling users to judge, trust, and question an AI effectively. Instead, the design of an interactive framework of multiple explanation methods was proposed, to achieve better user focus. Further, we would suggest setting up studies that do not just measure the user's classification performance on detecting true AI decisions but also measure if they can detect non-wanted biases or shortcuts \cite{geirhos2020shortcut}. A shortcoming of the proposed evaluation method is that it is still relatively costly to execute, being a user study requiring human involvement. We hope to see further human-centered XAI studies that extend our approach to further proxy tasks to create a solid ground for future XAI research and the creation of trustworthy AI applications.

\subsubsection{Disclosure of Interests}
The authors declare no conflict of interest.

\newpage

\bibliographystyle{splncs04}
\bibliography{bibliography}

\newpage
\appendix
\pagenumbering{roman}
\renewcommand\thefigure{\thesection.\arabic{figure}}
\setcounter{figure}{0} 

\begin{subappendices}
\renewcommand{\thesection}{\Alph{section}}%

\section{Additional Visualizations}

\begin{figure}[H]
\centering
\begin{subfigure}{1\textwidth}
    \centering
    \includegraphics[width=10cm,height=10cm,keepaspectratio]{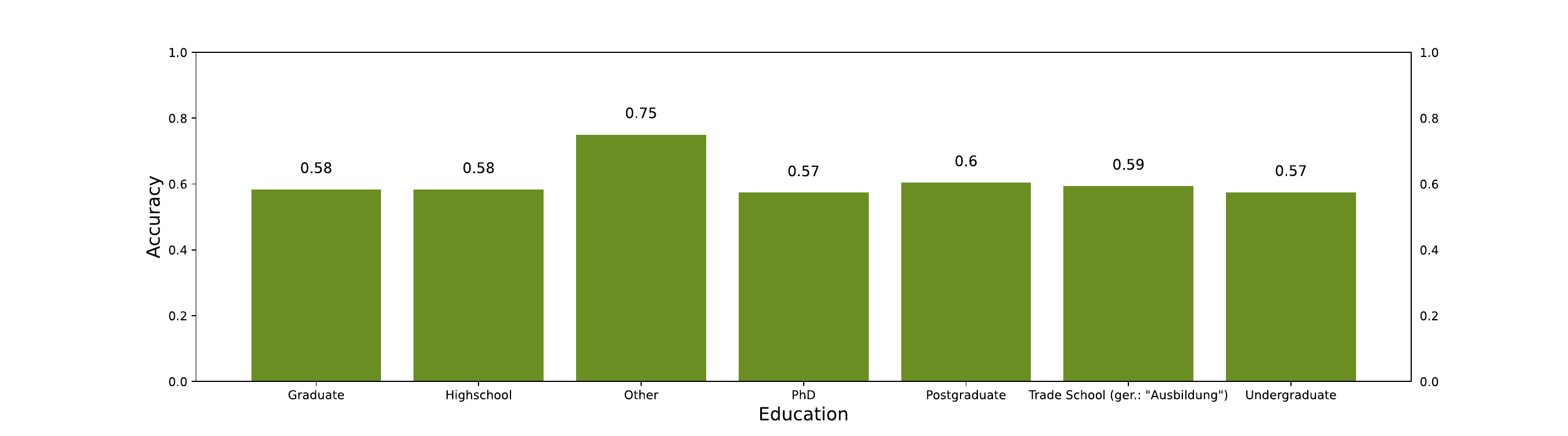}
    \label{fig:res_education}
    \caption{educational background}
\end{subfigure}
\begin{subfigure}{1\textwidth}
    \centering
    \includegraphics[width=10cm,height=10cm,keepaspectratio]{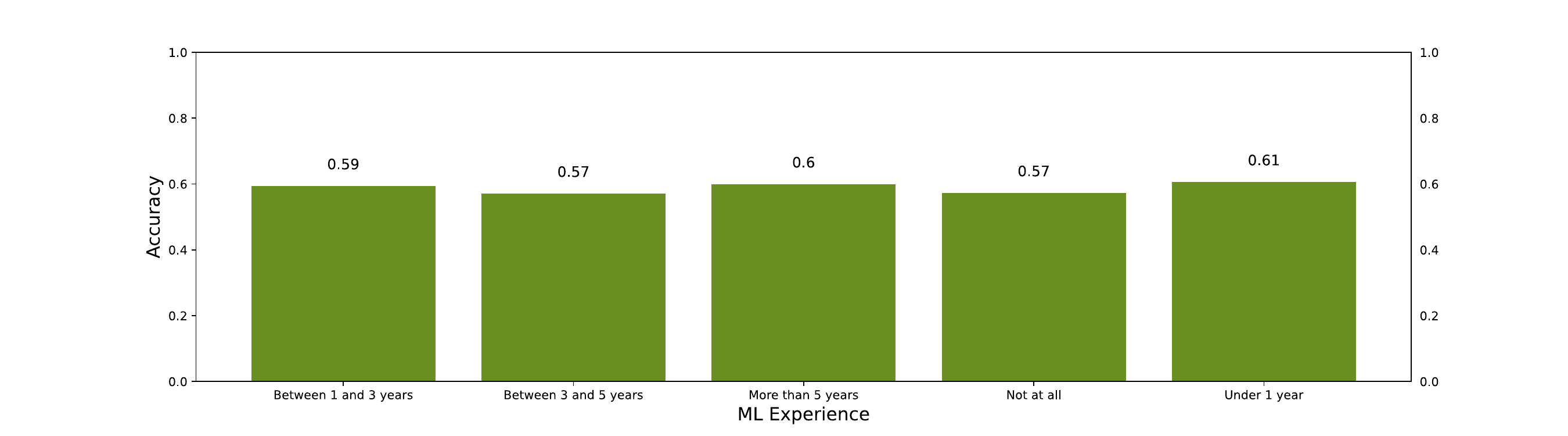}
    \label{fig:res_exper_years}
    \caption{machine learning experience in years}
\end{subfigure}
\begin{subfigure}{1\textwidth}
    \centering
    \includegraphics[width=10cm,height=10cm,keepaspectratio]{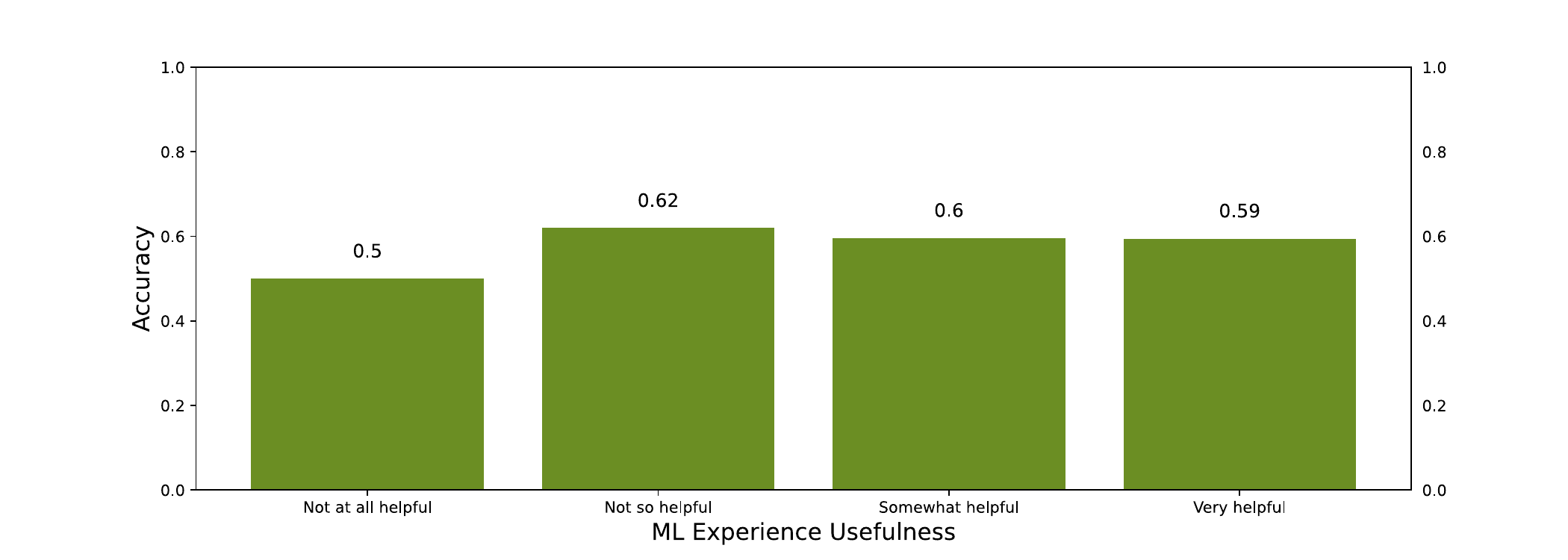}
    \label{fig:res_exper}
    \caption{self-assessment of the usefulness of machine learning experience for answering the survey}
\end{subfigure}
\begin{subfigure}{1\textwidth}
    \centering
    \includegraphics[width=10cm,height=10cm,keepaspectratio]{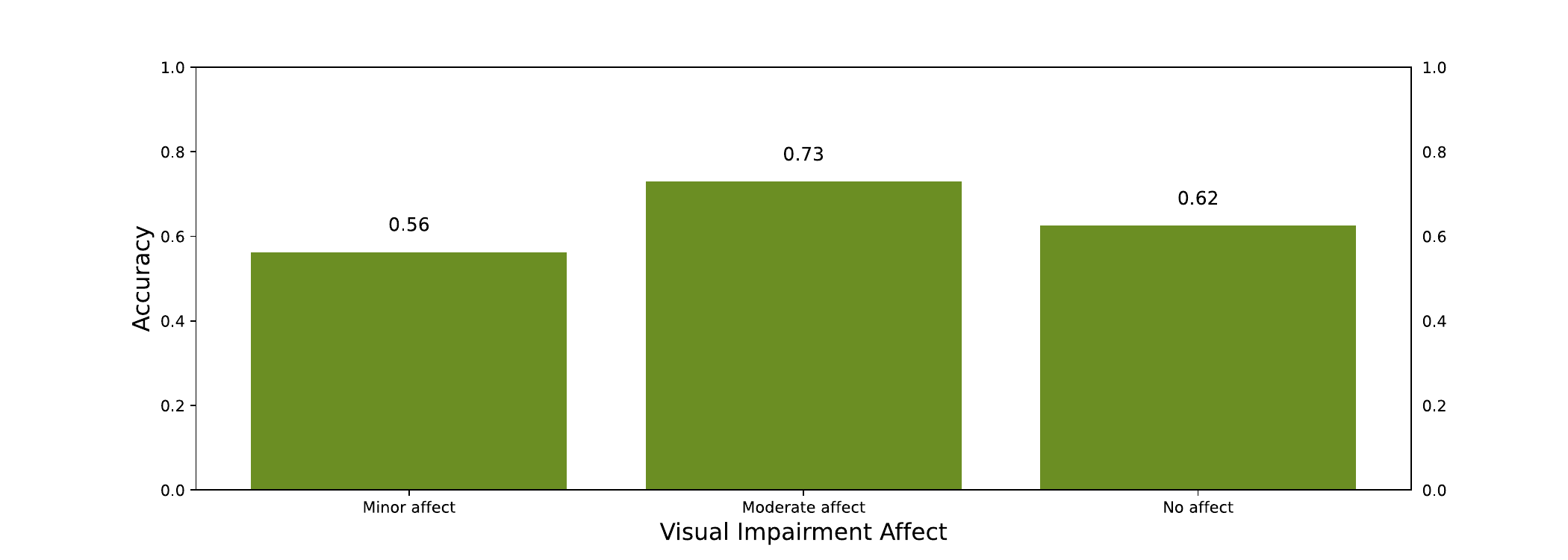}
    \label{fig:res_exper}
    \caption{self-assessment on visual impairment}
\end{subfigure}
\caption{Mean Accuracy based on educational background, machine learning experience in years,  self-assessment of the usefulness of machine learning experience, and self-assessment on visual impairment.}
\label{fig:demo}
\end{figure}

\begin{figure}[H]
\centering
\includegraphics[width=1\textwidth]{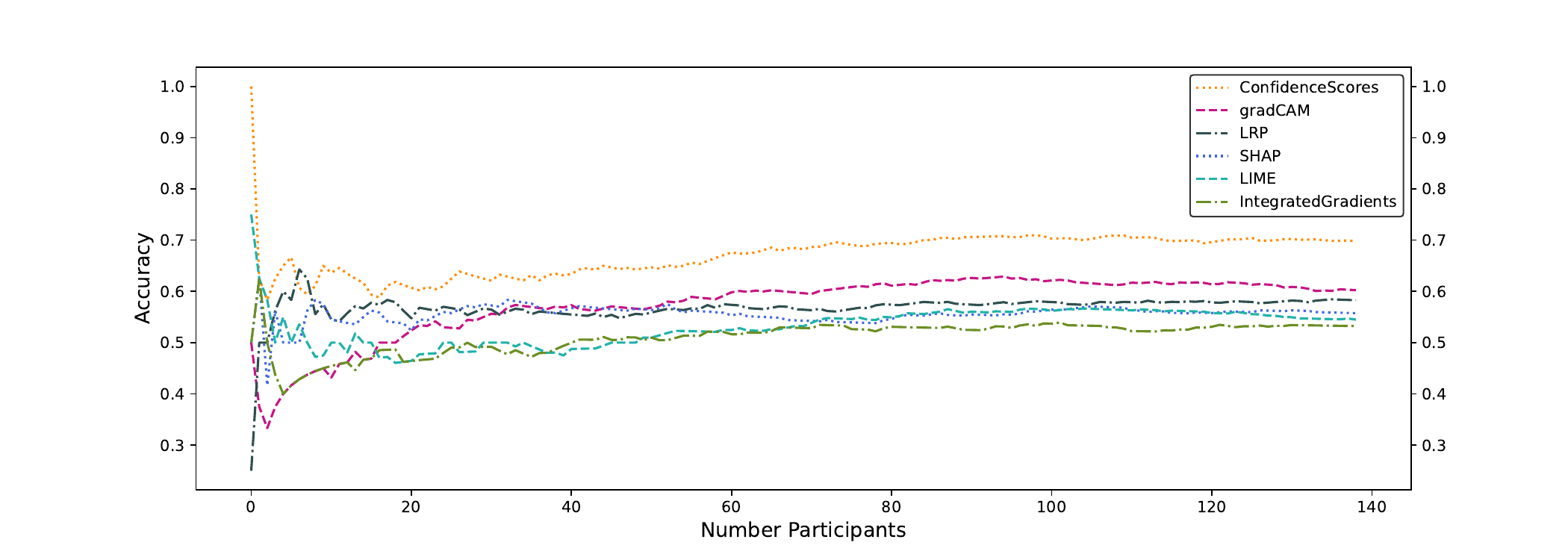}
\caption{Convergence of accuracy over the number of participants.} \label{fig:res_coner}
\end{figure}

\begin{figure}[H]
\centering
\includegraphics[width=1\textwidth]{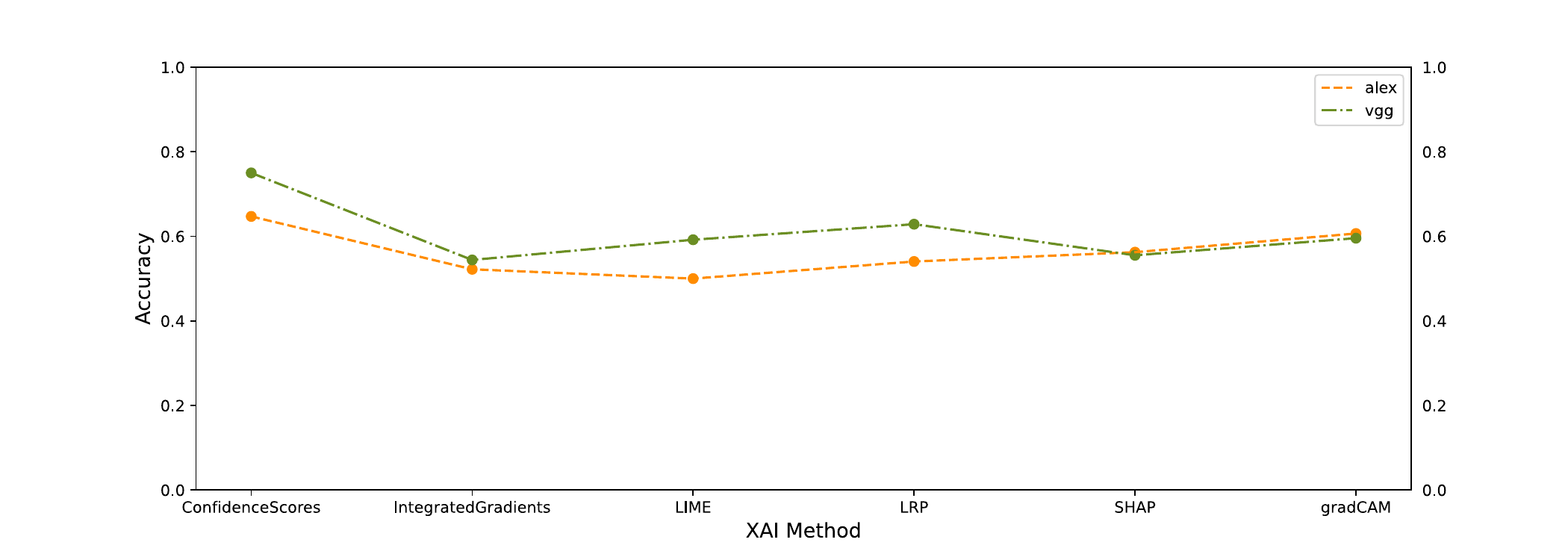}
\caption{Difference in accuracy between VGG16 and AlexNet}
\label{fig:res_models}
\end{figure}

\begin{figure}[H]
\centering
\includegraphics[width=0.99\textwidth]{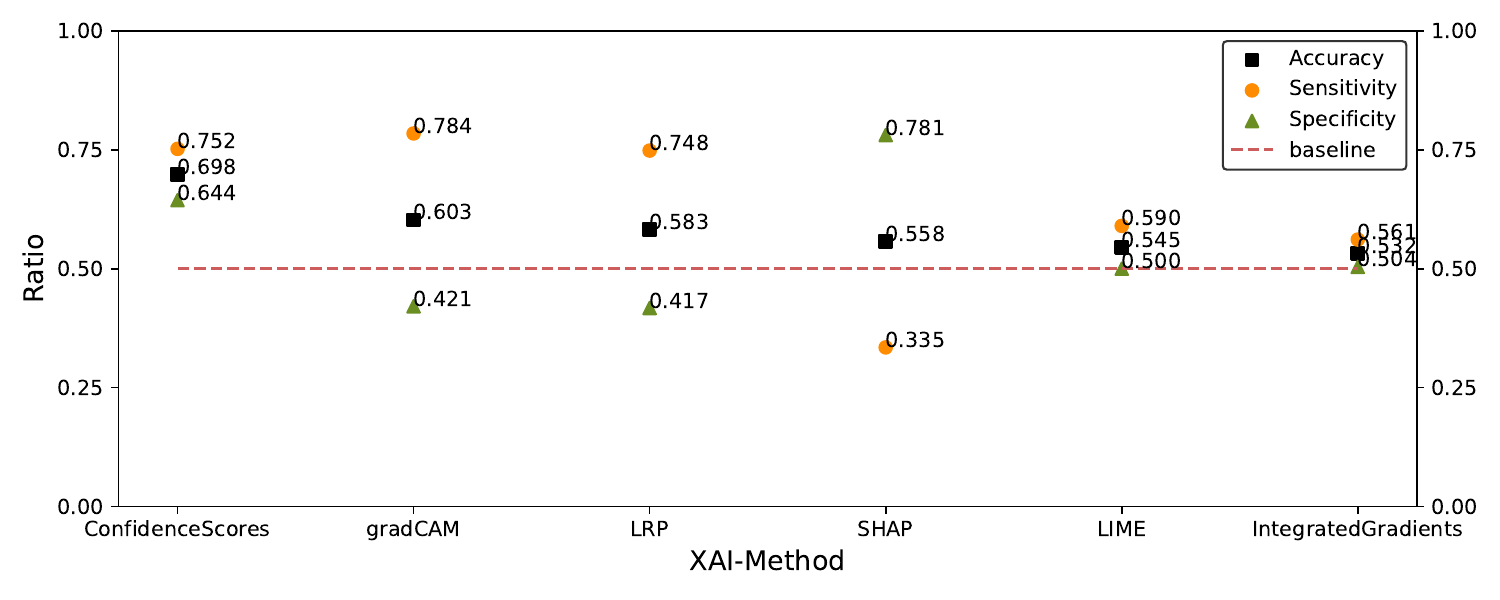}
\caption{Summary of means for accuracy, sensitivity, and specificity for all examined XAI methods compared to the random baseline.} \label{fig:metrics_overview}
\end{figure}

 
\renewcommand\thefigure{\thesection.\arabic{table}}
\setcounter{table}{0}  
\section{Demographic overview of participants}
\begin{table}[h]

\caption{Results of demographic questions}
\label{Demographic_table}

\centering
\begin{adjustbox}{width=1\textwidth}
\begin{tabular}{ccccccc}
\hline
\multicolumn{7}{|c|}{\textbf{Education}}                                                                                                                                                                                                                                         \\ \hline
\multicolumn{1}{|c|}{High School} & \multicolumn{1}{c|}{Trade School} & \multicolumn{1}{c|}{Undergraduate} & \multicolumn{1}{c|}{Graduate}           & \multicolumn{1}{c|}{Post Graduate}   & \multicolumn{1}{c|}{PhD}              & \multicolumn{1}{c|}{Other}         \\ \hline
\multicolumn{1}{|c|}{20}         & \multicolumn{1}{c|}{4}            & \multicolumn{1}{c|}{24}            & \multicolumn{1}{c|}{44}                 & \multicolumn{1}{c|}{26}              & \multicolumn{1}{c|}{7}                & \multicolumn{1}{c|}{1}             \\ \hline
                                 &                                   &                                    &                                         &                                      &                                       &                                    \\ \cline{3-7} 
                                 & \multicolumn{1}{c|}{}             & \multicolumn{5}{c|}{\textbf{ML Experience}}                                                                                                                                                               \\ \cline{3-7} 
                                 & \multicolumn{1}{c|}{}             & \multicolumn{1}{c|}{None}          & \multicolumn{1}{c|}{Less than 1 year}   & \multicolumn{1}{c|}{Between 1 and 3} & \multicolumn{1}{c|}{Between 3 and 5}  & \multicolumn{1}{c|}{More than 5}   \\ \cline{3-7} 
                                 & \multicolumn{1}{c|}{}             & \multicolumn{1}{c|}{65}            & \multicolumn{1}{c|}{33}                 & \multicolumn{1}{c|}{26}              & \multicolumn{1}{c|}{7}                & \multicolumn{1}{c|}{5}             \\ \cline{3-7} 
                                 &                                   &                                    &                                         &                                      &                                       &                                    \\ \cline{4-7} 
                                 &                                   & \multicolumn{1}{c|}{}              & \multicolumn{4}{c|}{\textbf{Perceived helpfulness of ML experience}}                                                                                                 \\ \cline{4-7} 
                                 &                                   & \multicolumn{1}{c|}{}              & \multicolumn{1}{c|}{Not at all helpful} & \multicolumn{1}{c|}{Not so helpful} & \multicolumn{1}{c|}{Somewhat helpful} & \multicolumn{1}{c|}{Very helpful} \\ \cline{4-7} 
                                 &                                   & \multicolumn{1}{c|}{}              & \multicolumn{1}{c|}{2}                  & \multicolumn{1}{c|}{15}              & \multicolumn{1}{c|}{43}               & \multicolumn{1}{c|}{11}            \\ \cline{4-7} 

\end{tabular}
\end{adjustbox}
\end{table}
\end{subappendices}
\end{document}